\documentclass[pre,aps,amsmath,amssymb,superscriptaddress,notitlepage,twocolumn]{revtex4-2}
\usepackage[colorlinks,bookmarks=false,citecolor=blue,linkcolor=blue,urlcolor=blue]{hyperref}
\usepackage[all]{hypcap}   
\usepackage{placeins}

\usepackage{amsmath,amssymb}
\usepackage{graphicx}
\graphicspath{{figures/}{/}}
\usepackage{dcolumn}
\usepackage{bm}
\usepackage[capitalise,compress]{cleveref}
\crefname{section}{Sec.}{Secs.}
\Crefname{section}{Section}{Sections}

\usepackage{verbatim}
\usepackage{color}
\usepackage{soul}
\usepackage{placeins}  
\usepackage{flafter}     
\usepackage{color}
\usepackage{hhline}
\usepackage{physics}
\usepackage{hyperref}

\usepackage{float}
\usepackage{caption}
\usepackage{subcaption}
\usepackage{lipsum} 

\newcommand{\Krylov}{\mathcal{K}}

\newcommand{\liouv}{\mathcal{L}}
\newcommand{\obs}{\mathcal{O}}

\usepackage[mathscr]{eucal}
\usepackage{tikz}
\usepackage{xcolor}

\newcommand{\bo}{\begin{outline}}
\newcommand{\eo}{\end{outline}}

\newcommand{\qed}{\nobreak \ifvmode \relax \else
      \ifdim\lastskip<1.5em \hskip-\lastskip
      \hskip1.5em plus0em minus0.5em \fi \nobreak
      \vrule height0.75em width0.5em depth0.25em\fi}
\begin{document} 

\title{Dependence of Krylov complexity saturation on the initial operator and state}

\author{Sreeram PG}
\email{sreerampg7@gmail.com}
\affiliation{Department of Physics, Indian Institute of Science Education and Research, Pune 411008, India}
\author{J. Bharathi Kannan}
\email{bharathikannan1130@gmail.com}
\affiliation{Department of Physics, Indian Institute of Science Education and Research, Pune 411008, India}
\author{Ranjan Modak}
\email{ranjan@iittp.ac.in}
\affiliation{ Department of Physics, Indian Institute of Technology Tirupati, Tirupati,  517619, India} 
\author{S. Aravinda}
\email{aravinda@iittp.ac.in}    
\affiliation{ Department of Physics, Indian Institute of Technology Tirupati, Tirupati,  517619, India} 


\begin{abstract}
Krylov complexity, a quantum complexity measure which uniquely characterizes the spread of a quantum state or an operator, has recently been studied in the context of quantum chaos. However,  the definitiveness of this measure as a chaos quantifier is in question in light of its strong dependence on the initial condition. This article clarifies the connection between the Krylov complexity dynamics and the initial operator or state. We find that the saturation value of Krylov complexity depends monotonically on the inverse participation ratio (IPR) of the initial condition in the eigenbasis of the Hamiltonian. We explain the reversal of the complexity saturation levels observed in \href{https://doi.org/10.1103/PhysRevE.107.024217}{ Phys. Rev. E 107, 024217 (2023)}  using the initial spread of the operator in the Hamiltonian eigenbasis. 
IPR dependence is present even in the fully chaotic regime, where popular quantifiers of chaos, such as out-of-time-ordered correlators and entanglement generation, show similar behaviour regardless of the initial condition. Krylov complexity averaged over many initial conditions still does not characterize chaos.

\end{abstract}

\maketitle
How an isolated quantum system thermalizes is closely connected to the scrambling of an initially localized, far-from-equilibrium quantum state. 
As the quantum state scrambles, one expects it to become more \textit{complex} in an intuitive sense. However, when trying to quantify this complexity rigorously, one soon encounters the ambiguity of the basis. For example, the complexity of a computer circuit would depend on the fixed set of gates available. Things look rather grim for a quantum state as the number of basis sets is infinite. Once a basis is fixed, the spread complexity can be defined as a function of the state's amplitude in the orthogonal basis directions, capturing the extent of the state. Krylov basis provides a unique solution to the ambiguity by minimizing the spread complexity over all the possible bases \cite{balasubramanian2022quantum}.
Krylov complexity maps the complexity of a quantum state to the average position of a particle hopping on a one-dimensional chain, with the probability of hopping determined by \textit{Lanczos} coefficients.

Since the spreading of quantum information in the system is a key feature of quantum chaos, Krylov complexity seems to be a suitable candidate for quantifying chaos in such systems. Similar to the other popular quantum chaos measures such as out-of-time-ordered correlator (OTOC) and entanglement generation, Krylov complexity also displays a ramp followed by saturation. The complexity growth rate is exponential in chaotic systems and the Krylov exponent upper bounds the Lyapunov exponent \cite{parker2019universal,nandy2024quantum}. Studies showed that the late-time saturation value of Krylov complexity carries signatures of chaos \cite{rabinovici2022krylov,rabinovici2022krylov1, nizami2023krylov, nizami2024spread}. Integrable systems tend to saturate at lower complexity values compared to nonintegrable systems. The difference in saturation is attributable to the difference in the variance of the Lanczos coefficients \cite{rabinovici2022krylov,rabinovici2022krylov1,rabinovici2021operator, nizami2023krylov, nizami2024spread}.

However, the legitimacy of Krylov complexity as a measure of quantum chaos has recently been questioned. The saturation value of the operator complexity can be completely changed by choosing a different initial operator or state \cite{espanol2023assessing,scialchi2024integrability}.
Another study in conformal field theory models \cite{dymarsky2021krylov} pointed out that the Krylov complexity can increase exponentially even when the system remains noninteracting. Similar false indication of chaos is also present in nonchaotic systems with saddle points \cite{bhattacharjee2022krylov}. Furthermore, the Krylov exponent keeps increasing monotonically in some  systems where the Lyapunov exponent is non-monotonic \cite{chapman2025krylov}, prompting the authors to conjecture that the Krylov complexity works more as an entropy and less as a measure of chaos. 

This conflict in the literature raises the question: What constitutes a good initial operator (or a state) for the Krylov
complexity study if there is one at all? Using traceless operators as suggested in \cite{rabinovici2022krylov} does not always work, as the counter-example in \cite{espanol2023assessing} points out. Another work \cite{scialchi2024integrability} on state complexity finds that both well-delocalized and sufficiently localized initial states in the energy eigenbasis can lead to wrong conclusions about chaos in the system. If we have to scrabble around to find \textit{suitable} operators and states so that Krylov complexity behaves \textit{as expected}, can it even be trusted to diagnose chaos? The key to unlocking this problem is to understand how Krylov complexity depends on the initial condition. 
Based on the initial spread in the Hamiltonian eigenbasis, we find a surprisingly simple relationship between the complexity dynamics and the initial operators and states. 

We study Krylov saturation of both states and operators in time-independent and periodically driven Floquet systems. For generality, we investigate a random matrix model for the unitary evolution of two coupled systems, with a transition to chaos depending on the coupling strength. After establishing the relationship between complexity and the initial condition in this model, we examine the well-studied quantum kicked top and transverse field Ising model, where we highlight the contrast between Krylov complexity and other widely accepted measures of quantum chaos. Ultimately, we investigate whether Krylov saturation can still be used to quantify chaos, despite its dependence on the initial state.
In this paper, Krylov complexity will refer to either states or operators, as clarified by context.
 

\textit{Krylov complexity:} Given a Hamiltonian $H$ of dimension $d$, a Hermitian operator $\obs$, and the Liouvillian superoperator $\liouv = \comm{H}{\cdot}$, the Krylov subspace is defined as the minimum subspace of $\liouv$ that contains $\obs(t)$ at all times, that is, $\Krylov =\{|\obs),\,\liouv|\obs),\,\liouv^2|\obs), \dots\}$, where $|\obs)$ is the state representation of $\obs$ in the Hilbert space of the operator. Using the Lanczos or Arnoldi iteration method \cite{parlett1998symmetric, viswanath1994recursion,sahu2023quantifying}, the set $\Krylov$ can be orthogonalized to obtain the Krylov basis vectors $\{ |K_n)\}_{n=1}^{d_\Krylov},$ where $d_\Krylov$ denote the dimension of the Krylov subspace. In this paper, we use the Arnoldi method for numerical simulations. The Krylov complexity of $\mathcal{O}(t)$   is defined as
\begin{equation}
    K_\mathcal{C}(t)= \sum_{n=0}^{d_k-1} n \left| ({K_n}| {\mathcal{O}(t)}) \right|^2. 
\end{equation}
A similar definition follows for the state complexity by simply replacing the operator $\obs$ by a state $\ket{\psi}$. The extension of the definition to include periodically driven systems can be found in \cite{yates2021strong,nizami2023krylov}, and also in the supplementary file.

\textit{Inverse Participation Ratio for states and operators}
Let  the Hamiltonian $H$ or the Floquet $\mathcal{U}$ for a periodically driven system  has eigenvectors $\{|v_i\rangle\}_{i=1}^{d}$. For an initial state $|\psi\rangle$, the IPR with respect to this eigenbasis is:
\begin{equation}
    \mathrm{IPR}_{\mathcal{U}}(|\psi\rangle)= \sum_{i=1}^{d} |\langle v_i|\psi\rangle|^4, \label{eq ipr_state}
\end{equation}
IPR varies between $\frac{1}{d}$ and one, equal to one when $\ket{\psi}$ is an eigenstate of the Floquet, and $\frac{1}{d}$ when $\ket{\psi}$ is perfectly  delocalized in $\{|v_i\rangle\}.$ 

Similarly, we define IPR for an operator $\mathcal{O}.$ To treat all operators on the same footing, first, we normalize $\mathcal{O}$ by defining $ \mathcal{O}=\mathcal{O}/\mathrm{Tr} \sqrt{\mathcal{O}\mathcal{O}^\dagger}.$ 
 Expanding $\mathcal{O}$ in the energy eigenbasis obtained above, we get $\mathcal{O}= \sum_{i,j=1}^d \bra{v_i}\mathcal{O} |v_j\rangle  \ket{v_i}\bra{v_j} .$  We define the IPR of $\mathcal{O}$ using the diagonal elements in this representation as follows:
 \begin{equation}
    \mathrm{IPR}_{H}(\mathcal{O})= \sum_{i=1}^d {|\bra{v_i}\mathcal{O}\ket{v_i}|^2}. \label{ipr_op}
\end{equation}
 Equation (\ref{ipr_op}) is a function of the operator's projections along the energy eigenvectors of the dynamics. It quantifies the initial overlap and the spread of $\mathcal{O}$ along the eigenvector directions. One must note that $\sum_{i=1}^d |\bra{v_i}\mathcal{O}\ket{v_i}| \neq 1,$ unlike the case of states. $\mathcal{O}$ could as well be a hollow matrix in this representation, with diagonal entries all zero if $\mathcal{O}$ has zero overlap with the one-dimensional projectors along $\{\ket{v_i}\}.$

\textit{Krylov state complexity with random matrix transition ensemble dynamics: }
To examine $K_\mathcal{C}(t)$ in the context of quantum chaos, we choose a dynamics that transitions from regular to chaotic as a function of a parameter. To be independent of any particular model system, we consider a random matrix model of interacting bipartite systems with Floquet dynamics over one period of the form (also known as the Random matrix transition ensemble (RMTE) \cite{srivastava2016universal}),
\begin{equation}
    \mathcal{U}_\epsilon= U_{12}(\epsilon)( U_1 \otimes U_2) \label{unitary}.
\end{equation}
Here, the unitaries $U_1$ and $U_2$ are chosen from the circular unitary ensemble (CUE), and they act on the  Hilbert spaces of subsystem-1  and subsystem-2 respectively, each of dimension $d$. The parameter $\epsilon$ governs the strength of the interaction between the two subsystems in $U_{12}(\epsilon).$ The coupling unitary is diagonal, with the non-zero elements being $\mathrm{exp}(i 2\pi \epsilon \mathscr{\xi}_{n_1n_2})$, where $1 \leq n_1,n_2 \leq d,$ and $\xi_{n_1n_2} \in [-\frac{1}{2}, \frac{1}{2})$, chosen uniformly at random. At $\epsilon=0,$ the two subsystems are non-interacting. As $\epsilon$ increases from zero to one, the eigenphase spacing distribution of the unitary matrix transitions from Poissonian to Wigner-Dyson statistics. 
\begin{figure*}[!t]
    \centering
    \includegraphics[height=1.8in,width=\linewidth]{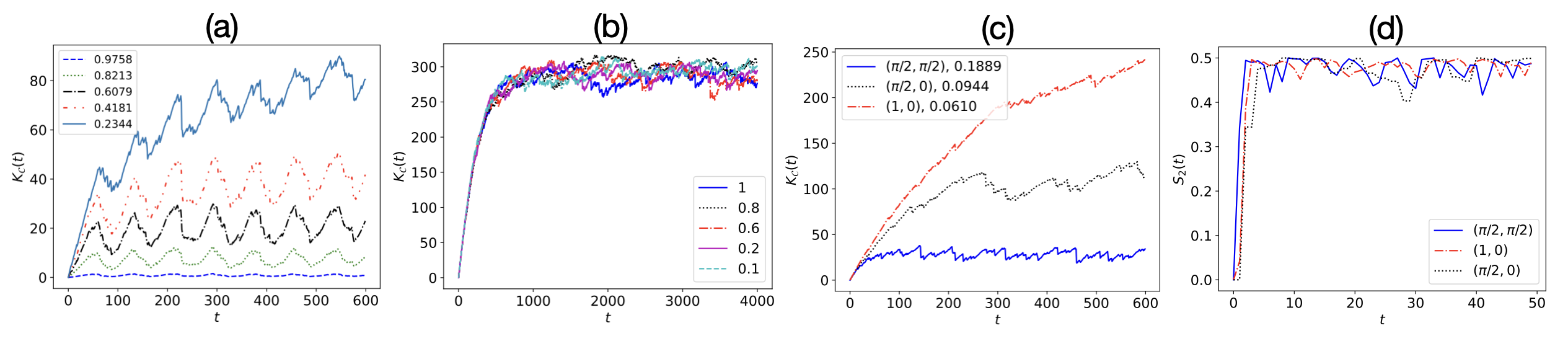}
    \caption{(a) Krylov complexity as a function of time for initial states with different IPR in the eigenbasis of $\mathcal{U}_{\epsilon=1}$. The dimensions of the subspaces are chosen to be $d=5.$ There is a monotonic relationship between the IPR of the state and the rate of growth and saturation of the complexity. (b)$K_C(t)$ for the states with the same IPR under different RMTE dynamics. Legends denote the  $\epsilon$ values of $\mathcal{U}_\epsilon$.  (c)$K_C(t)$ of three spin coherent states with $(\theta, \phi)$  under the kicked top dynamics. The numeral outside the parentheses in the legend is the IPR. The kick-strength $\kappa=6,$ and $j=15.$  (d) Linear entropy evolution for the coherent initial states in (c) under the same unitary dynamics.}
    \label{fig:kc_ipr_state}
\end{figure*}

First, we fix the evolution given by Eq. (\ref{unitary}), by letting $d=5$, and choosing an $\epsilon.$ Let us choose $\epsilon=1$ so that the evolution is chaotic and the evolution unitary is denoted by $\mathcal{U}_{\epsilon=1}.$ The set of random numbers $\{\mathscr{\xi}_{n_1n_2} \}$ once generated, is fixed for the entire experiment.

The next step is to pick initial states as inputs to the Floquet evolution. We choose states with different IPRs in the eigenbasis of $\mathcal{U}_{\epsilon=1}.$  To obtain such states, we  rotate $\ket{v_1}$, an eigenvector of $\mathcal{U}_{\epsilon=1}$ using a unitary rotation operator 

\begin{equation}
R(\theta,\phi)=\exp[i \theta ( j_x \mathrm{sin}(\phi)- j_y \mathrm{cos}(\phi))] \label{eq:rotation}
\end{equation}for different $(\theta,\phi).$ Here 
$j_x$ and $j_y$ are angular momentum operators acting in the collective Hilbert space of $\mathcal{U}_{\epsilon=1}.$  
  The $K_\mathcal{C}(t)$ for states with different IPRs is shown in Fig. \ref{fig:kc_ipr_state}(a). Note that the system is in the chaotic limit, yet the saturation of $K_\mathcal{C}(t)$ depends on the IPR of the initial state.  If the state has $\mathrm{IPR}=1,$ it does not evolve, and the Krylov complexity remains zero. The more spread the initial state is in $\{|v_i\rangle\}$, the higher the growth rate and saturation of its complexity. Even when the number of Krylov basis vectors is the same for two states, their complexity could differ based on how the states spread within the Krylov space.

Now that the behavior of the states with different IPR under a chosen dynamics is clear, we probe the converse situation. We now vary $\epsilon$ so that the dynamics determined by $\mathcal{U}_\epsilon$ are different. But we fix the initial state to be the maximally delocalized state for all $\mathcal{U}_\epsilon$ so that the IPR are the same. More explicitly,  when subsystems are of $d=5,$ the initial state has equal superposition of all the eigenstates:
$\ket{\psi_0}=\frac{1}{\sqrt(25)}(\ket{v_1}+\ket{v_2}+...\ket{v_{25}}).$ Here the $\{\ket{v_i}\}$ vary depending on the $\mathcal{U_\epsilon},$ but the IPR is the same and is equal to $0.04$ regardless of $\epsilon.$  We find that the $K_\mathcal{C}(t)$, irrespective of the $\epsilon$ rise and saturate very close to each other, as shown in Fig.\ref{fig:kc_ipr_state}(b). Since there are fluctuations post-saturation, we plot a long-time evolution for clarity. We also studied other states more localized in $\{\ket{v_i}\}$, maintaining the same IPR.  Their behavior was the same. Figure \ref{fig:kc_ipr_state}(b) is a clear indication that the Krylov complexity is independent of the nature of the dynamics and only depends on the IPR. 

How does the correlation between the Krylov saturation and the initial spread of the state come about?  It turns out that the $K_C(t)$ of a state $\ket{\psi_0}$ at large $t$ can be expressed as (details in the supplementary):
\begin{equation}
    K_\mathcal{C}(t)= \sum_j \left(\sum_{i}i|\bra{K_i}v_j\rangle |^2\right) |\bra{v_j}\psi_0\rangle |^2,
\end{equation}
where $\{\ket{v_j}\}$ are the eigenstates of the Hamiltonian. When the overlap  $|\bra{v_j}\psi_0\rangle|^2$ is larger, the overlap of $\ket{E_j}$ with the orthogonal Hilbert subspace $(\{\bra{K_i}v_j\rangle |^2\}_{i>0}$ is smaller and vice versa.  Moreover, a small change in $|\langle v_j\ket{\psi_0}|^2$ gets magnified in the opposite sense in  $\left(\sum_{i}i|\bra{K_i}v_j\rangle |^2\right) $, owing to the integer weight factors. Therefore, to get a larger Krylov complexity, it is best for $|\langle v_j\ket{\psi_0}|^2$ to be as small as possible. However, since  $\sum_j|\langle v_j\ket{\psi_0}|^2=1$, not all of them can be simultaneously small. The best scenario is $\langle v_j\ket{\psi_0}|^2=1/d, \: \forall j,$ which corresponds to the lowest IPR state.
A more detailed analysis is given in the supplementary. 

\textit{Krylov state complexity with the quantum kicked top:} Turning to a physical system that is well-studied in the context of quantum chaos, we find that $K_C(t)$ can be a misleading measure of chaos. The quantum kicked top is a periodically driven system, with the evolution for one period given by the unitary \cite{haake1987classical,Haake2018}
\begin{equation}
    U = \exp (-i \frac{\kappa}{2j} J_z^2 ) \exp(-i \alpha  J_y). \label{eq: top}
\end{equation}
Here $\kappa$ is the kick strength, determining the amount of chaos in the system.  $\alpha$ is the angle of precession about $y$-axis, which we fix to be  $\pi/2,$  and $j$ is  the spin angular momentum.

To study $K_\mathcal{C}(t),$ we choose $j=15,$ and $\kappa=6.$ At this $\kappa,$ chaos dominates
the corresponding classical phase space \cite{haake1987classical}. The complexity evolution for three different initial spin coherent states is shown in Fig. \ref{fig:kc_ipr_state}(c), again revealing the monotonic dependence on IPR.  It should be noted that we are witnessing IPR dependence even in this chaotic regime. 

On the other hand, studying the entanglement dynamics of the kicked top considered as a compound system of spins \cite{ghose2004entanglement} shows a contrasting picture. The linear entropy  of a single spin with the rest increases and rapidly saturates close to the maximum entropy of $0.5,$ at $\kappa=6.$ The linear entropy is defined as $S_2(t)= 1- \mathrm{Tr} \rho_s^2(t),$ where $\rho_s(t)$ is the reduced density matrix of a single spin. The $S_2(t)$ evolution is shown in Fig. \ref{fig:kc_ipr_state}(d) for the same initial states as in Fig. \ref{fig:kc_ipr_state}(c); however, it is now impossible to differentiate one initial state from the other based on evolution. This behavior persists for random $(\theta, \phi)$ initializations as expected from an underlying chaotic phase space.

\begin{figure}
    \centering
    \includegraphics[width=0.8\linewidth]{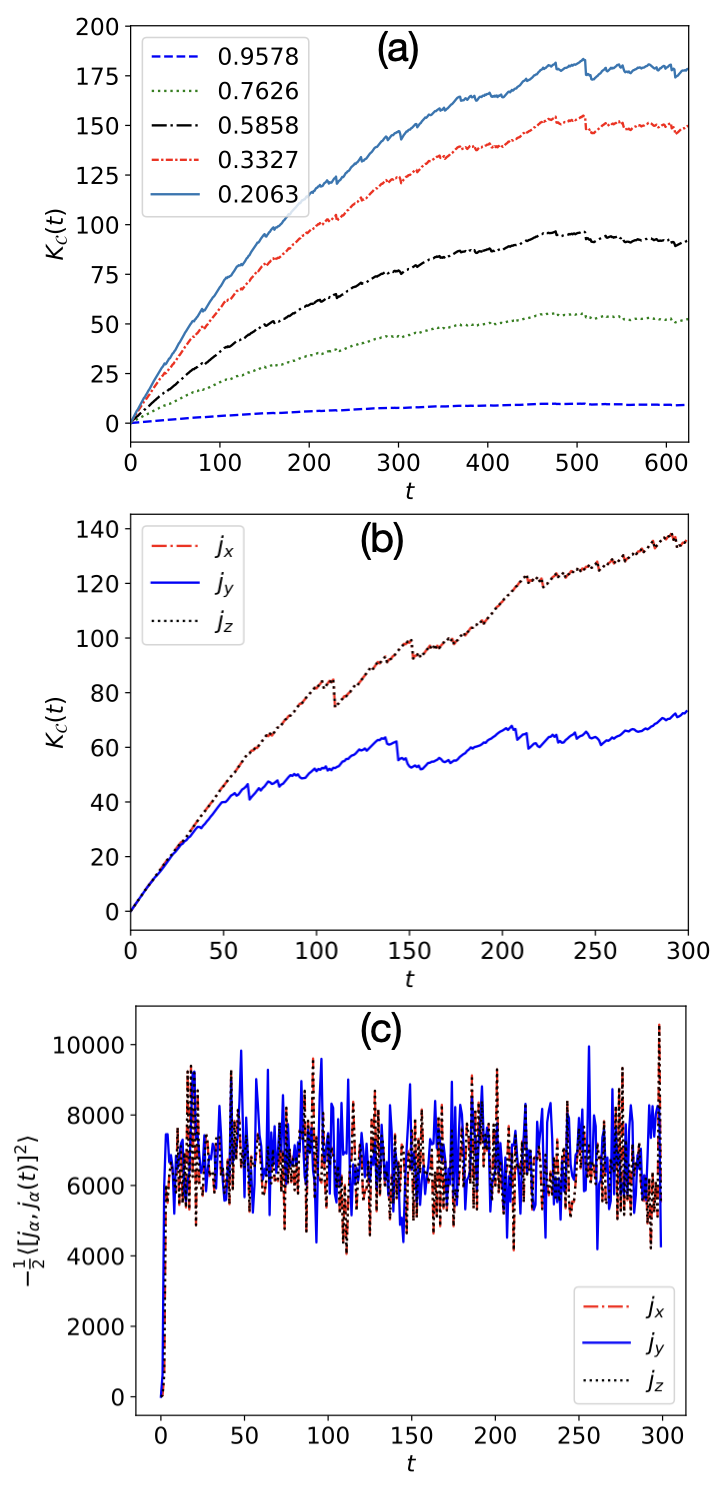}
    \caption{(a) Operator complexity for initial operators with different IPR in the eigenbasis of $\mathcal{U}_{\epsilon=1}$ of the RMTE. (b) Complexity evolution of spin operators under kicked top unitary with $j=15$ and $\kappa=6.$ Krylov complexity of $j_x$ and $j_z$ are the same, whereas that of $j_y$ is lower. The corresponding IPRs obtained are $j_x=j_z=0$, $j_y=0.009$. An inverse correlation between IPR and $K_C(t)$ is evident from the figure. (c) OTOC of the form $-\frac{1}{2} \langle [j_\alpha,j_\alpha(t)]^2 \rangle$, where $\alpha \in \{x,y,z\}$ show that OTOC cannot distinguish these operators. OTOC expectation is taken with respect to the maximally mixed state. }
    \label{fig: operator_complexity}
\end{figure}

\textit{Krylov operator complexity with random matrix transition ensemble dynamics:}
We first study the IPR dependence of the operator complexity for the random matrix evolution in Eq. (\ref{unitary}). We fix $\epsilon=1$ as in the case of states so that we have a chaotic evolution governed by $\mathcal{U}_{\epsilon=1}$.  To generate initial operators with different IPR, we rotate $\mathcal{U}_{\epsilon=1}$ using the rotation operator $R(\theta,\phi)$ in Eq. (\ref{eq:rotation}) by varying $(\theta,\phi).$  The resultant dynamics of these operators under $\mathcal{U}_{\epsilon=1}$ is shown in Fig. \ref{fig: operator_complexity}(a). The figure displays the dependence of the Krylov complexity on the IPR  of the initial operator.

\textit{Krylov operator complexity with the quantum kicked top:}  We illustrate another example of $K_\mathcal{C}(t)$ not capturing chaos and instead following the IPR in the quantum kicked top with $j=15$ and $\kappa=6.$ 
Figure \ref{fig: operator_complexity}(b) shows the complexity evolution under Eq. (\ref{eq: top}) with angular momentum operators $(j_x,j_y,j_z)$ as initializations. The figure shows that in the chaotic regime, the evolutions of $j_x$ and $j_z$ are identical and differ from $j_y$. This is in line with the IPR of these operators. Both $j_x$ and $j_z$ are hollow matrices in the energy eigenbasis and have vanishing IPR according to Eq. (\ref{ipr_op}). Krylov complexity evolution of both these operators is therefore identical as seen in Fig. \ref{fig: operator_complexity}(b). $j_y$ has nonzero IPR (0.009), and hence lower complexity.  

However, OTOC evolution of the form $-\frac{1}{2} \langle [j_\alpha,j_\alpha(t)]^2 \rangle$, where $\alpha \in \{x,y,z\}$ cannot resolve between these operators, as  Fig. \ref{fig: operator_complexity}(c) shows.
In a fully chaotic regime, OTOC and entanglement dynamics do not depend on the initial condition, as true chaos measures should. With the phase space lacking regular structures, one point is no different from another. However, $K_C(t)$ shows different evolutions based on the IPR even in this regime,  indicating a serious defect.

\textit{Krylov operator complexity with transverse field Ising model: } 
The authors in \cite{espanol2023assessing}, working with a transverse field Ising model, found that the saturation level of  $K_C(t)$ changes based on the initial operator. The Ising Hamiltonian has nearest-neighbour coupling and a transverse magnetic field in the $X-Z$ plane, given by:
\begin{equation}
    H= \sum_{k=1}^L (h_x \sigma_k^x +h_z \sigma_k^z) - J \sum_{k=1}^{L-1} \sigma_k^z \sigma_{k+1}^z. \label{eq: tfim}
\end{equation}
Here, $L$ is the chain length, $J$ is the interaction strength and  $h_x$ and $h_z$ denote the components of the magnetic field. The operators $\sigma_k^\alpha,$ where $\alpha \in \{x,y,z\}$ are local Pauli operators.  Fixing $J= h_x =1$, the system transitions progressively from integrability to chaos as $h_z$ decreases from $2.5$ to $0.2.$

A stark display of the effect of the initial condition on $K_C(t)$ is obtained by choosing $S_z =\sum_{k=1}^{L} \sigma_k^z$ and $S_x =\sum_{k=1}^{L} \sigma_k^x$ as initial operators. The system has a reflection symmetry about the centre of the chain, and the operators $S_x$ and $S_z$ lie in the positive parity sector. Working in the corresponding symmetry subspace,  the complexity for $S_z$ evolves and saturates higher in the chaotic regime compared to the regular region (see \cite{espanol2023assessing} or supplementary for the figure). This is in line with the findings in \cite{rabinovici2022krylov, rabinovici2022krylov1}, and is consistent with what one would expect from studying other measures of chaos.
The initial choice of $S_x =\sum_{k=1}^{L} \sigma_k^x$ makes  $K_C(t)$ behave in a way opposite to this expectation, and now there is more complexity in the integrable regime (see \cite{espanol2023assessing} or supplementary for the figure)!

We explain this behaviour using the spread of the initial operator in the eigenbasis of the transverse-field Ising Hamiltonian. The IPR for the operators $S_x$ and $S_z$ according to Eq. (\ref{ipr_op}) is shown in the table below. In each row, the operator with the smaller IPR consistently shows a higher $K_C(t)$ saturation. 
\begin{center}
\begin{tabular}{ |c|c|c| } 
 \hline
 $h_z$ & IPR for $S_x$ & IPR for $S_z$ \\
 \hline
0.2 & $0.3545$  & $0.0946$ \\
\hline
1.35  & $0.2307$  & $0.3391$ \\
 \hline
2.5 & $0.1351$&$0.4409$ \\ 
 \hline
\end{tabular}
\end{center}

\textit{Average $K_C(t)$:} Given the dependence on the initial condition, $K_\mathcal{C}(t)$ is unreliable in characterising chaos with a single initial condition. However, can it at least roughly estimate chaos when averaged over initial conditions? 
\begin{figure*}[hbt!]
    \centering
    \includegraphics[width=1\linewidth]{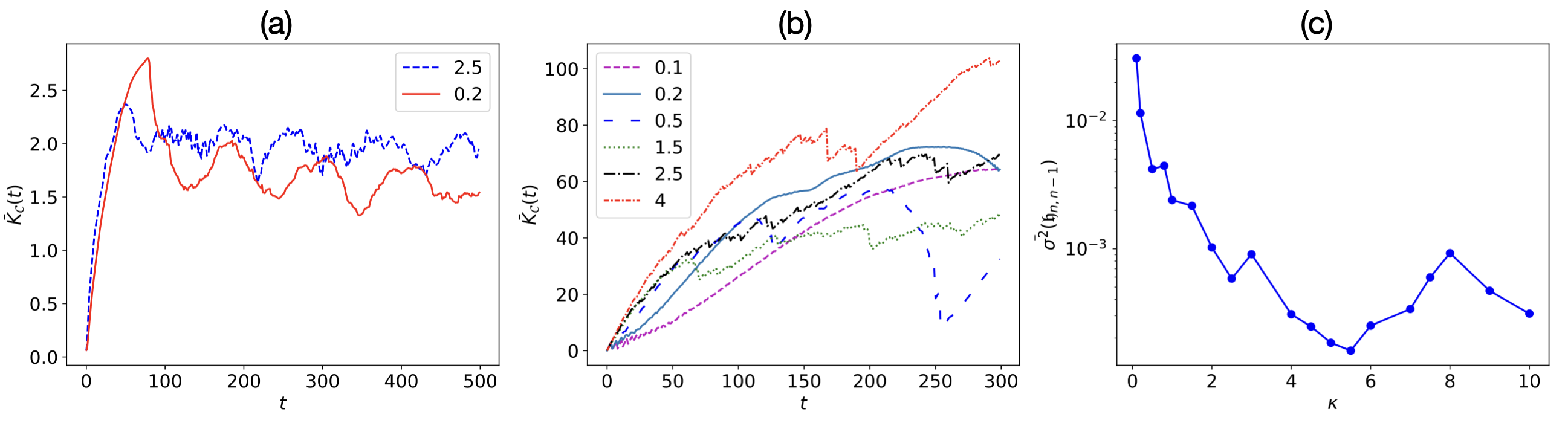}
    \caption{(a) Average Krylov complexity $\bar{K}_\mathcal{C}(t)$ for the transverse Ising model with $L=6$ over 100 initial operators, as mentioned in the main text. (b) $\bar{K}_\mathcal{C}(t)$ averaged over 100 randomly chosen initial coherent states under kicked top evolution with $j=10$ for different kick strengths $\kappa.$ (c) Variance of  Arnoldi coefficients $\mathfrak{h}_{n,n-1}$  against kicking strength obtained after averaging over 100 initial coherent state evolutions using kicked top with $j=10$.}
   \label{fig:average kc}
\end{figure*}

We study the transverse field Ising model in Eq. (\ref{eq: tfim}), averaging over one hundred initial operators in the positive parity subspace.
To generate the initial operators, we rotate each spin -$\frac{1}{2}$ sites in $S_x =\sum_{k=1}^{L} \sigma_k^x$ using the rotation operator $R(\theta,\phi)=\exp[i \theta ( \sigma_x \mathrm{sin}(\phi)- \sigma_y \mathrm{cos}(\phi))]$ for $(\theta, \phi)$ chosen uniformly at random, such that $\theta \in [0,\pi)$ and $\phi \in [0,2\pi)$.
We plot the average Krylov complexity (denoted $\bar{K}_\mathcal{C}(t)$) for $h_z=0.2$ corresponding to the chaotic regime, and $h_z=2.5$ in the regular regime, in Fig. \ref{fig:average kc}(a). Other parameters of the model are kept the same as before, namely $L=6,\:\& \: J=h_x=1.$ Complexity at $h_z=0.2$ does not saturate above that of $h_z=2.5,$ even with the averaging. Therefore, Krylov complexity does not describe chaos even as a coarse measure.
Similarly, $\bar{K}_\mathcal{C}(t)$ for one hundred spin coherent states evolved with the kicked top Floquet in Fig \ref{fig:average kc} (b) shows no monotonic correlation between complexity and the amount of chaos.

The only part of the Krylov construction, that somewhat correlates with chaos is the variance of the Arnoldi coefficients denoted by $\mathfrak{h}_{n,n-1}$ (see \cite{nizami2023krylov}, or Krylov construction for Floquet systems in the supplementary), shown in Fig. \ref{fig:average kc}(c). Although we have averaged over one hundred initial coherent states in Fig. \ref{fig:average kc}(c), this behavior remains robust even for single-state evolution. A similar observation about the dispersion of the Lanczos coefficients being a better chaos diagnostic can be found in \cite{espanol2023assessing,scialchi2024integrability,hashimoto2023krylov}. 
However, the initial drop in this quantity is sharp, followed by a gradual decrease to saturation, with notable fluctuations in the chaotic regime. We do not consider it an accurate measure of chaos due to its non-monotonicity.

\textit{Discussion:} This article clarifies that the  $K_C(t)$ saturation depends on the IPR of the initial condition and not on the amount of chaos in the system. Although the  Krylov basis is generated from the system dynamics, the associated complexity fails to capture the information scrambling caused by the same dynamics. Other well-accepted measures, such as entanglement and OTOC, are based on quantum correlations that are physical and basis-independent. 
While the Krylov basis is unique for each initial condition, the complexity is still a basis-dependent quantity. States and operators with complexity saturation uncorrelated with chaos are common, making the average Krylov complexity an ineffective chaos measure.

Someone might argue that OTOC can also show a false signal of chaos \cite{hashimoto2020exponential,xu2020does} in nonchaotic systems, if calculated in the neighborhood of a saddle point. However, the physical reason for this behavior is evident and can be mapped back to the underlying classical phase space. Furthermore, this perceived problem is solved by a long-time evolution of the  OTOC \cite{kidd2021saddle}, which shows strong oscillations, unlike the chaotic regime. Using logarithmic OTOC \cite{trunin2023quantum,trunin2023refined} is another solution. 
Similarly, initial-state dependence is present with entanglement generation in the mixed-phase space regime \cite{lombardi2011entanglement}, which can be resolved by averaging over initial states \cite{madhok2015comment}. In the fully chaotic regime, such ambiguities disappear and averaging is unnecessary \cite{lombardi2015reply}. In contrast, $K_C(t)$ remains ambiguous even in fully chaotic regime.

Krylov saturation is not a measure of chaos; what about other aspects of Krylov complexity? For instance, the variance of the Arnoldi coefficients is linked to chaos, as is the height of the Krylov peak before saturation \cite{balasubramanian2022quantum,baggioli2025krylov,camargo2024spread,huh2024spread,erdmenger2023universal,alishahiha2024krylov}. While the former shows reasonable correlation with chaos, the peak is not universal \cite{camargo2024spread,alishahiha2024} and requires large Hilbert space to manifest \cite{balasubramanian2022quantum, baggioli2025krylov}.
The limited states analysed in the literature do not confirm if the Krylov peak reliably indicates chaos when saturation fails. Exploring its validity as a chaos signature could be a valuable future direction.

\textit{Acknowledgments:} S PG acknowledges the I-HUB quantum technology foundation (I-HUB QTF), IISER Pune, for financial support. Authors thank Prof. M.S. Santhanam for useful discussions.

\FloatBarrier
\bibliography{ref}
\bibliographystyle{elsarticle-num}

\end{document}


\title{%
  {\red{Supplementary Information}} \\
  \large Dependence of Krylov complexity saturation on the seed operator and state}

\author{Sreeram PG}
\email{sreerampg7@gmail.com}
\affiliation{Department of Physics, Indian Institute of Science Education and Research, Pune 411008, India}
\author{J. Bharathi Kannan}
\email{bharathikannan1130@gmail.com}
\affiliation{Department of Physics, Indian Institute of Science Education and Research, Pune 411008, India}
\author{Ranjan Modak}
\email{ranjan@iittp.ac.in}
\affiliation{ Department of Physics, Indian Institute of Technology Tirupati, Tirupati,  517619, India} 
\author{S. Aravinda}
\email{aravinda@iittp.ac.in}    
\affiliation{ Department of Physics, Indian Institute of Technology Tirupati, Tirupati,  517619, India} 
\maketitle
\section{Krylov construction}
Given a Hamiltonian $H$, a Hermitian operator $\obs$, and the Liouvillian superoperator $\liouv = \comm{H}{\cdot}$, the Krylov subspace is defined as the minimum subspace of $\liouv$ that contains $\obs(t)$ at all times, that is, $\Krylov =\{|\obs),\,\liouv|\obs),\,\liouv^2|\obs), \dots\}$, where $|\obs)$ is the state representation of $\obs$ in the operator's Hilbert space.

Given an inner product $(\obs_1|\obs_2) = \Tr\,(\obs_1^\dag\,\obs_2)$, the iterative Lanczos algorithm can be used to generate an orthonormal basis of this subspace. The exact form of this algorithm consists of the following steps \cite{parker2019universal, rabinovici2021operator,parlett1998symmetric,viswanath1994recursion}:

 \textbf{Lanczos algorithm:}

\begin{itemize}
	\item Define auxiliary variables: \\ $b_0 = 0$, $~|\obs_{-1}) = 0$.
	\item Normalize the operator to expand:\\
		$|\obs_0) = |\obs) / \left(\obs|\obs\right)^{\frac{1}{2}}$.
	\item for $n = 1, 2, \cdots$, repeat:
		\begin{itemize}
			\item $|{\mathcal{A}}_n) = \liouv\,|\obs_{n-1}) - b_{n-1}\,|\obs_{n-2})$.
			\item $b_n = \left(\mathcal{A}_n|\mathcal{A}_n\right)^{\frac{1}{2}}$. If $b_n = 0$, stop.
			\item $|\mathcal{O}_n) = |\mathcal{A}_n) / b_n$
		\end{itemize}
\end{itemize}

\textbf{Full orthogonalization (Arnoldi iteration):}

To avoid numerical instability associated with the above algorithm, explicit orthogonalization of the new vector with all the previous ones can be performed at each step. The steps are as follows \cite{parlett1998symmetric,rabinovici2021operator}.

\begin{itemize}
    \item $|\obs_0) = |\obs) / \left(\obs|\obs\right)^{\frac{1}{2}}$, $b_0=\left(\obs|\obs\right)^{\frac{1}{2}} $.
    \item for $n\geq 1$ ; $|{\mathcal{A}}_n) = \liouv\,|\obs_{n-1})$\\
    $|{\mathcal{A}}_n) \xrightarrow{} \liouv\,|\obs_{n-1}) - \sum_{m=0}^{n-1} |\obs_{m}) (\obs_m|\mathcal{A}_{n})$.
    \item Repeat the previous step once again, to ensure orthogonality.
    \item $b_n= {(\mathcal{A}_n|\mathcal{A}_n)}^\frac{1}{2},$ called Arnoldi coefficients.
    \item if $b_n=0,$ stop. Else, $|\obs_n)= \frac{1}{b_n}|\mathcal{A}_n)$
    
\end{itemize}

\section{Krylov construction for Floquet systems for Operators}

 This approach employs the Arnoldi iteration to systematically construct the Krylov basis in a periodically driven quantum system. Operator complexity for studying operator growth is defined similarly. Let $\mathcal{O}_0$ be the operator under study at $t=0$ and construct \cite{nizami2023krylov, nizami2024spread}
 
\begin{align}   
\mathcal{H}_K^O&= \{|\mathcal{O}_0), |U_F^{\dagger}\mathcal{O}_0 U_F),   |(U_F^{\dagger})^ 2 \mathcal{O}_0 U_F^ 2),....\} \nonumber\\
&=\{|\mathcal{O}_0), |\mathcal{O}_1), |\mathcal{O}_2), .... \}\, 
\end{align} 
where $\mathcal{O}_j$ denotes the Heisenberg picture operator at time $t=jT$. $|\mathcal{O})$ denotes the operator $\mathcal{O}$ as a ket in the linear space of all operators that act in $\mathcal{H}$. Note that here $U_F$, as a superoperator, has the action $U_F|\mathcal{O})=|U_F^ {\dagger} \mathcal{O} U_F)$. The Krylov basis is then generated by the following recursive algorithm. Define $|K_0)=\mathcal{O}_0)$ and

\begin{align}  
&|K_1)=\frac{1}{\mathfrak{h}_{1,0}}\big[ U_F|K_0)-\mathfrak{h}_{0,0}|K_0) \big] \\ 
&|K_n)=\frac{1}{\mathfrak{h}_{n,n-1}}\left[ U_F |K_{n-1})-\sum_{j=0}^ {n-1}\mathfrak{h}_{j,n-1}|K_j) \right] \label{Kn}
\end{align}  

with $\mathfrak{h}_{j,k}=(K_j|U_F|K_k)=\dfrac{1}{D}Tr(K_j^{\dagger} U_F^{\dagger}K_{k}U_F )$. The normalisation $\mathfrak{h}_{n,n-1}$ are the Arnoldi coefficients. With $\mathcal{O}_j=(U_F^{\dagger})^j \mathcal{O}_0U_F^ j$ being the time evolved operator at (stroboscopic) time $t=jT$, the operator complexity, defined by $\mathcal{K}_j^{O}=(\mathcal{O}_j|\hat{K}|\mathcal{O}_j)$, is given by

\begin{align}
\label{KCFdefn}
\mathcal{K}_j^{O}&=\sum_{n=0}^{D_K-1} n |(K_n|U_F^ j|K_0)|^2 \nonumber\\
&= \frac{1}{D^ 2}\sum_{n=0}^{D_K-1} n |Tr[K_n^{\dagger} U_F^{\dagger j} K_0 U_F^j]|^2.
\end{align}

This gives us a direct way to compute operator complexity given the Floquet matrix and the Krylov operator basis. Using $|\mathcal{O}_{j})=\sum_n \phi_n^j |K_n)$, where $\phi_n^j=(K_n|\mathcal{O}_j)$ is the $n$-th operator amplitude at time $t=j T$, we can write an equivalent form of the above equation

\begin{equation}
\label{KCFdefn2}
\mathcal{K}_j^{\mathcal{O}}=\sum_n  n \, |\phi_n^j|^2.
\end{equation}

\section{Level spacing ratio}
\begin{figure}
    \centering
    \includegraphics[width=0.5\textwidth]{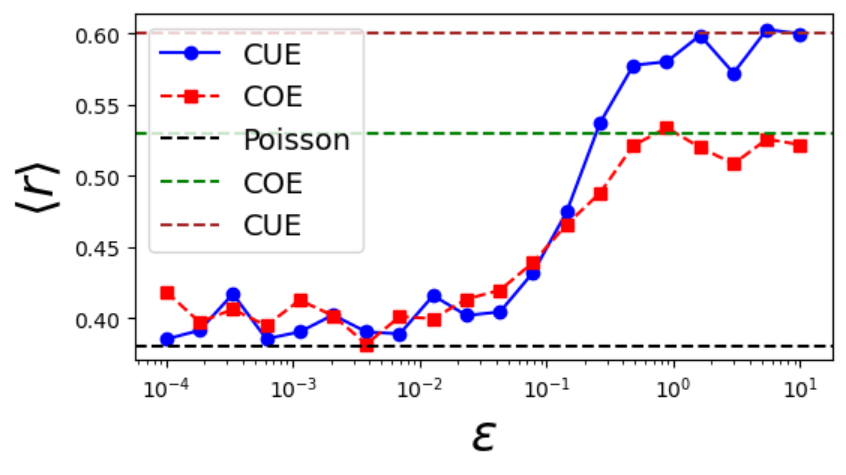}
    \caption{Level spacing ratio of RMTE model}
    \label{fig:level_spacing}
\end{figure}

Spectral properties of the system can distinguish between integrable and chaotic systems \cite{bohigas1984characterization,gutzwiller2013chaos}. The ratio of the consecutive eigenvalues \cite{atas2013distribution} for the random matrix transition ensemble described in the main text is plotted in Fig. \ref{fig:level_spacing}. As $\epsilon$ varies, the system transitions from a regular to chaotic regime, as indicated by the ratio distributions.
The level spacing ratio is defined as:
\begin{equation}
    r_n = \frac{\min(s_n, s_{n+1})}{\max(s_n, s_{n+1})},
\end{equation}
where $s_n = E_{n+1} - E_n$ represents the spacing between consecutive eigenvalues (eigenphase in case of floquet unitary). For integrable systems, the eigenvalues tend to follow a Poisson distribution, leading to an average spacing ratio of $\langle r \rangle \approx 0.386$. In contrast, for chaotic quantum systems that follow random matrix theory statistics. For instance, systems exhibiting time-reversal symmetry and belonging to the COE(circular orthogonal ensemble) have an average ratio of $\langle r \rangle \approx 0.530$. In contrast, those in the CUE (circular unitary ensemble) have $\langle r \rangle \approx 0.599$.

\section{Krylov complexity in Transverse field Ising model}
\begin{figure}[!t]
    \centering
    \begin{subfigure}[t]{0.45\textwidth}
        \centering
        \includegraphics[height=2.2in]{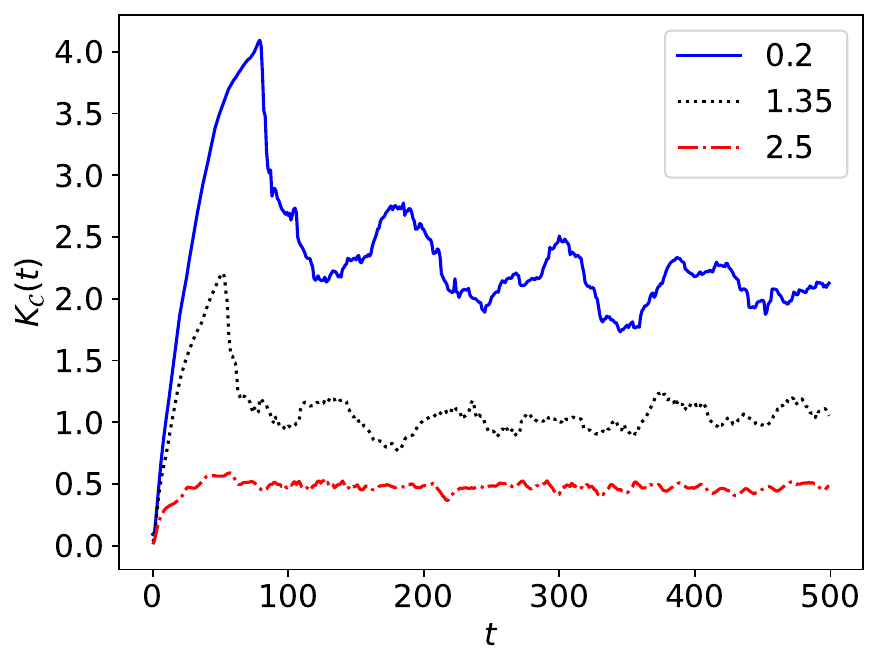}
        \caption{}
    \end{subfigure}%
    \hfill
    \begin{subfigure}[t]{0.45\textwidth}
        \centering
        \includegraphics[height=2.2in]{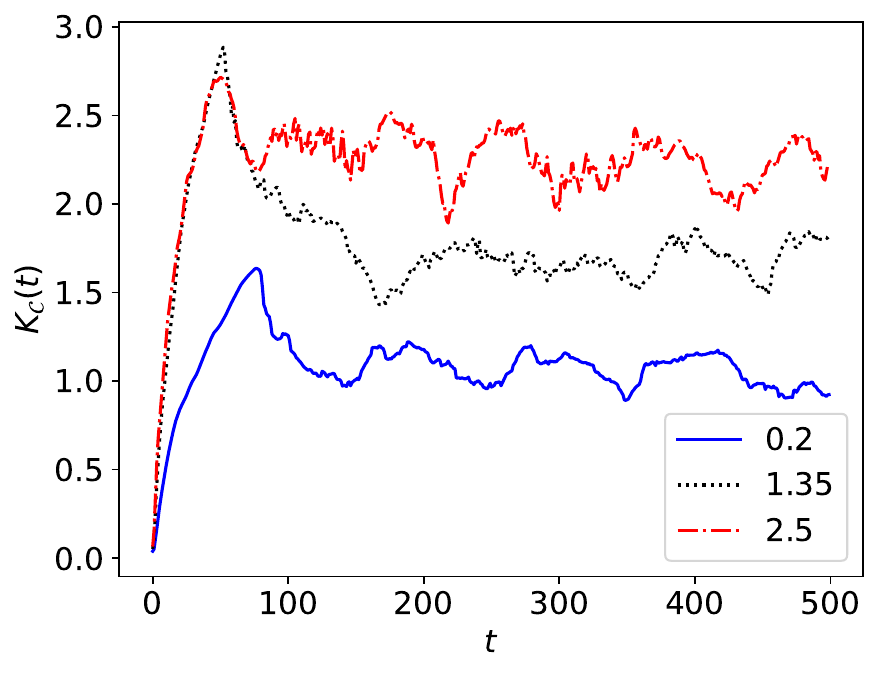}
        \caption{ }
    \end{subfigure}
    \caption{ The saturation levels of Krylov complexity flip depending on the initial operator in the transverse Ising model. We choose a spin length $L=6,$ and  $J= h_x =1$ in the Hamiltonian.  In (a), the initial operator is $S_z$, whereas in (b) the initial operator is $S_x$. }
    \label{fig: operator_complexity}
\end{figure}

The Ising Hamiltonian includes a nearest-neighbour coupling and a transverse magnetic field in the $X-Z$ plane, given by:
\begin{equation}
    H= \sum_{k=1}^L (h_x \sigma_k^x +h_z \sigma_k^z) - J \sum_{k=1}^{L-1} \sigma_k^z \sigma_{k+1}^z. \label{eq: tfim}
\end{equation}
Here, $L$ is the chain length, $J$ is the interaction strength and  $h_x$ and $h_z$ denote the components of the magnetic field. The operators $\sigma_k^\alpha$ are local Pauli  Fixing $J= h_x =1$, the system transitions progressively from integrability to chaos as $h_z$ decreases from $2.5$ to $0.2.$ The flip in the saturation of complexity depending on the initial operator is shown in Fig. \ref{fig: operator_complexity}.

\section{Dependence of Krylov complexity on IPR}

Let the initial state of the system be denoted by $\ket{\psi_0}$, and the unitary determining the dynamics, obtained from the Hamiltonian $H$, be denoted by $U.$ In its eigenbasis, $U$ has the following decomposition:
\begin{equation}
    U= \sum_j \exp(i \phi_j) \ket{E_j} \bra{E_j},
\end{equation}
where $\{\ket{E_j}\}$ are the eigenvectors of $U.$
After $n-$ action of the unitary,
\begin{align}
    \ket{\psi_n}&=  U^n \ket{\psi_0} \notag\\
    &=\sum_j \exp(in \phi_j) \bra{E_j}{\psi_0}\rangle\ket{E_j} \label{psin}
\end{align}
The Krylov complexity of $\ket{\psi_0}$ at time $t=n$, is given by 
\begin{equation}
    K_\mathcal{C}(n)= \sum_{i} i |\bra{K_i}\psi_n\rangle|^2, \label{krylov_def}
\end{equation}
where $\{\ket{K_i}\}_{i}$ are the orthogonal Krylov basis vectors, generated one at a time until $\ket{\psi_n}$ is spanned at all times. Substituting Eq.  \ref{psin} in Eq. \ref{krylov_def}, we get
\begin{align}
    K_\mathcal{C}(n)= & \sum_{i} i \sum_j |\exp(in\phi_j) \bra{K_i}E_j\rangle \bra{E_j}\psi_0\rangle |^2 \notag\\
    =&  \sum_{i} i \sum_j |\exp(in\phi_j)|^2 |\bra{K_i}E_j\rangle |^2|\bra{E_j}\psi_0\rangle |^2 \notag\\
    =&  \sum_{i} i \sum_j  |\bra{K_i}E_j\rangle |^2|\bra{E_j}\psi_0\rangle |^2 \notag\\
    =& \sum_j \left(\sum_{i}i|\bra{K_i}E_j\rangle |^2\right) |\bra{E_j}\psi_0\rangle |^2 \label{change_order},
\end{align}
where in Eq. \ref{change_order}, the order of the summation is flipped. Since we are interested in the saturated value of Krylov complexity, consider a time $n$ long enough so that all the Krylov vectors are already generated.
The term within the parentheses in Eq. \ref{change_order}
is a spread complexity of the state $\ket{E_j},$ in the Krylov space of $\ket{\psi_0}.$ The eigenstate $\ket{E_j}$ is assured to be spanned by $\{\ket{K_i}\}$, as long as the overlap $\bra{E_j}\psi_0\rangle \neq 0$,  since the Krylov space is the minimal invariant subspace containing $\ket{\psi_0}.$ Also, any $\ket{E_j} $ for which $\bra{E_j}\psi_0\rangle = 0$ is not spanned by the Krylov space, and their contribution to the Krylov complexity is zero.
Since  $\ket{K_0}=\ket{\psi_0}$, and  the coefficient $i$ multiplying $|\bra{K_0}E_j\rangle|^2$ is zero,
one can interpret the parentheses term as quantifying the spread of  $\ket{E_j}$ in the  Hilbert space orthogonal to the initial state  $\ket{\psi_0}$.

Note that $\{|\bra{E_j}\psi_0\rangle|^2\} \cup \{\bra{K_i}E_j\rangle |^2\}_{i>0}$ form a probability distribution, since $\ket{E_j}$ is normalized. Therefore, $|\bra{E_j}\psi_0\rangle|^2$ and $\{\bra{K_i}E_j\rangle |^2\}_{i>0} $ are mutually complementary. 
When the overlap  $|\bra{E_j}\psi_0\rangle|^2$ is larger, the overlap of $\ket{E_j}$ with the orthogonal Hilbert subspace $(\{\bra{K_i}E_j\rangle |^2\}_{i>0}$ is smaller and vice versa.  Moreover, a small change in $|\langle E_j\ket{\psi_0}|^2$ gets magnified in the opposite sense in  $\left(\sum_{i}i|\bra{K_i}E_j\rangle |^2\right) $, owing to the integer weight factors. For instance, suppose that $|\langle E_j\ket{\psi_0}|^2$ decreases by a small positive real number $\delta_0,$ for a different choice of $\ket{\psi_0}.$ Then correspondingly, the other probabilities change, producing a net increase of $\delta_0$, since probabilities sum to one. Denoting the changes $\delta \left(|\langle{K_i}\ket{E_j}|^2\right)= \delta_i, $ for $i>0, $ we have that $\delta_0= \sum_{i>0} \delta_i.$ Taking the weight factors into account, we get
\begin{equation}
     \sum_{i} i \delta_i \geq \delta_0.
\end{equation}
Thus, any small change $\delta_0$ in  $|\langle E_j\ket{\psi_0}|^2$ gets overcompensated by changes in $\left(\sum_{i}i|\bra{K_i}E_j\rangle |^2\right)$. 
Therefore, to get a larger Krylov complexity, it is best for $|\langle E_j\ket{\psi_0}|^2$ to be as small as possible. However, since  $\sum_j|\langle E_j\ket{\psi_0}|^2=1$, not all of them can be simultaneously small. The best scenario is $\langle E_j\ket{\psi_0}|^2=1/d, \: \forall j,$ where $d$ is the dimension of the Hilbert space. Any change from this state means clumping of the probability density to fewer, larger $|\langle E_j\ket{\psi_0}|^2$, which suppresses $\left(\sum_{i}i|\bra{K_i}E_j\rangle |^2\right)$ more than it benefits from larger $|\langle E_j\ket{\psi_0}|^2$. The minimum suppression of $\left(\sum_{i}i|\bra{K_i}E_j\rangle |^2\right)$ is achieved for the case of $\langle E_j\ket{\psi_0}|^2=1/d, \: \forall j$ where each terms contribute symmetrically to the Krylov complexity.

One can see the connection to state IPR straightaway from the above discussion. The condition $\langle E_j\ket{\psi_0}|^2=1/d, \: \forall j,$ corresponds to the most uniformly spread out state with the lowest IPR. Any increase in IPR skews the probability distribution, leading to a smaller Krylov complexity.
\subsection*{Proof of the optimality of the lowest IPR state}
 Beyond the intuitive arguments made above, one can rigorously prove that the state with $|\langle E_j | \psi_0\rangle|^2 \:\forall \: j$ having the highest IPR also gives the highest Krylov complexity. In Eq. \ref{change_order}, for simplicity of notations, let \( p_j = |\langle E_j | \psi_0\rangle|^2 \) and \( C_j = \sum_i i |\langle K_i | E_j\rangle|^2 \), so that
\begin{equation}
K_\mathcal{C}(n) = \sum_j C_j p_j.
\end{equation}
To find the point at which Krylov complexity is maximum, maximize $K_\mathcal{C}(n)$ subject to the constraint $\sum_j p_j=1,$ Define the Lagrangian:
\begin{equation}
\mathcal{L} = \sum_j {C}_jp_j - \lambda \left( \sum_j p_j - 1 \right).
\end{equation}
Note that $C_j$ are implicit functions of the probability distribution $(p_1,p_2...,p_d),$ where $d$ is the dimension of the Hilbert space. At a critical point: 
\begin{equation}
\frac{\partial \mathcal{L}}{\partial p_j} = C_j + \sum_k p_k \frac{\partial C_k }{\partial p_j}- \lambda = 0  \label{optimal}
\end{equation}
\begin{enumerate}
    \item \textbf{At the critical point, the derivatives $ \frac{\partial C_k }{\partial p_j}$ have to vanish:}

 Assume that $\frac{\partial C_k }{\partial p_j} \neq 0$. Then the weighted sum $\sum_k p_k \frac{\partial C_k}{\partial p_j}$ has to precisely cancel $C_j-\lambda$ for arbitrary $j$ at the critical point. The implicit dependence of $C_k$ on $p_j$  via the Krylov basis structure is a nontrivial relationship. Recall that the coefficients $C_k$ are determined by the structure of the Krylov vectors  ($C_k= \sum_i |\bra{K_i}E_k\rangle|^2$), and the Krylov vectors implicitly depend on $p_j.$ Since the Krylov vectors are generated by the initial state in conjunction with the  pre-determined Hamiltonian, there is no freedom to pick and choose $\{C_k\}$  so that $\sum_k p_k \frac{\partial C_k}{\partial p_j}$  precisely cancels $C_j-\lambda$ for arbitrary $j.$ 

Therefore, $\frac{\partial C_k }{\partial p_j}=0$ 
 and $ C_j=\lambda, \forall \: j$ at the critical point in Eq. \ref{optimal}.

\item
\textbf{Identical values of $C_j$ require special symmetry:}

A constant $C_j =\lambda \: \forall j$    satisfies the  critical point condition in Eq. \ref{optimal}, as argued above. However, identical values of $C_j$ require special symmetry in the overlaps $\bra{K_i}E_j\rangle.$  The value $C_j= \lambda \: \forall \: j$ is only possible if the following condition holds:
\begin{equation*} \textrm{\textbf{Condition1:}}\:|\bra{K_i}E_j\rangle|^2\: \textrm{are independent of} \:j \: \forall i. \label{condition}
\end{equation*} 
The above condition  indicates identical participation of $\ket{E_j}$ in the Krylov construction.  Since $\ket{K_0}=\ket{\psi_0}$, condition 1  implies equal projections $|\bra{E_j}\psi_0\rangle|^2=\frac{1}{d}$  ensuring the symmetry. Thus, the unique solution that satisfies the optimality condition in Eq. \ref{optimal} is $p_j=\frac{1}{d}\: \forall \:j.$  One can verify that the complexity value decreases in the neighborhood of the optimal solution by perturbing $\{p_j\}$ away from $1/d. $ Perturbation breaks the symmetry and leads to undue suppression of some $C_j.$ While some other  $C_j$ might benefit from the corresponding $p_j$ becoming smaller than $\frac{1}{d},$ their contribution gets down-weighted by the now smaller $p_j.$ Therefore, the net change in the complexity is negative.   Thus, $p_j=1/d \: \forall \:\: j $ must be at least a local maximum. 
\item{ \textbf{Uniqueness of the solution:} }

There are no zeros to Eq. \ref{optimal} other than $C_j=\lambda \: \forall \:j$, as $\sum_k p_k \frac{\partial C_k }{\partial p_j}$ has to arbitrarily adjust to get  zero as argued in point 1. $C_j=\lambda \: \forall \:j$  is only satisfied at  $p_j=1/d \: \forall\: j$. Any deviation from this symmetric point will asymmetrically change $C_j$ values, and they do not remain constant. Thus, there are no other solutions, and $p_j=1/d \: \forall\: j$ is also the global maximum.
\end{enumerate}
\section{Hamiltonian   variance and the Lanczos coefficients}
Apart from  IPR, the variance of the Hamiltonian with  the initial state also carries the information about the state-spread with respect to the dynamics. Hamiltonian takes a tridiagonal form, known as the Hessenberg form  in the Krylov basis \cite{balasubramanian2022quantum} . In terms of the Lanczos coefficients $\{a_n,b_n\},$ the Hamiltonian $(H)$ obeys the following equation:
\begin{equation}
    H\ket{{K_n}}=a_n\ket{K_n}+b_{n+1}\ket{K_{n+1}}+b_n \ket{K_{n-1}}, \label{krylov}
\end{equation}
where $\{\ket{K_n}\}$ denote the Krylov basis vectors.
Since the initial state $\ket{\psi_0}$ forms the first Krylov vector $\ket{K_0},$ the following equations stem out from Eq. \ref{krylov}.
\begin{align}
    H \ket{\psi_0} &= a_0 \ket{\psi_0} +b_1 \ket{K_1} \label{hamiltonian_action}\\
    H\ket{K_1} &=a_1\ket{K_1}+b_1\ket{\psi_0} +b_2\ket{K_2}.
\end{align}
Now it is straightforward to obtain the action of $H^2$ on the initial state $\ket{\psi_0}$ 
\begin{widetext}
   \begin{align}
    H^2\ket{\psi_0}&=H(a_0 \ket{\psi_0} +b_1 \ket{K_1}) \notag\\
    &=a_0(a_0 \ket{\psi_0} +b_1 \ket{K_1})+b_1(a_1\ket{K_1}+b_1\ket{\psi_0} +b_2\ket{K_2}) \notag \\
    &= a_0^2 \ket{\psi_0} +a_0b_1\ket{K_1}+ b_1a_1 \ket{K_1}+b_1^2\ket{\psi_0})+b_1b_2\ket{K_2}.
\end{align} 
\end{widetext}

Therefore,
\begin{equation}
    \bra{\psi_0}H^2\ket{\psi_0}=a_0^2+b_1^2 
\end{equation}
Also $\bra{\psi_0}H\ket{\psi_0}=a_0$ follows from Eq. \ref{hamiltonian_action}.
Hence, the variance $\Delta H^2$ with respect to the initial state is given by
\begin{align}
    \Delta H^2 &= \bra{\psi_0} H^2\ket{\psi_0} -\bra{\psi_0}H \ket{\psi_0} ^2 \notag\\
    &=a_0^2+b_1^2-a_0^2 \notag\\
    &=b_1^2.
\end{align}
Thus, $\Delta H^2$ depends only on a single Lanczos coefficient, namely the $b_1.$ This coefficient therefore contains the information about the spread of the initial state with respect to the Hamiltonian.
\FloatBarrier
\bibliography{ref}